\documentclass[12pt]{article}
\usepackage{ametsoc}

\begin{document}
\begin{titlepage}
\renewcommand{\baselinestretch}{1.5}
\begin{center}
\bfseries
\Large
Maximum Entropy Production and Non-Gaussian Climate Variability
\mdseries
\\[1cm]
\large
Philip Sura \\
{\em Department of Earth, Ocean and Atmospheric Science\\
The Florida State University, Tallahassee, Florida}
\\[3cm]
\today 
\\[6cm]
\normalsize
{\em Corresponding author address:}\\
Philip Sura\\
Department of Earth, Ocean and Atmospheric Science\\
The Florida State University\\
1017 Academic Way, Tallahassee, FL 32306-4520\\
Phone: (850) 644-1268, Fax:(850) 644-9642 \\
Email: psura@fsu.edu
\end{center}
\end{titlepage}


\renewcommand{\baselinestretch}{2}
\small\normalsize
\setlength{\parskip}{1.5ex}


\begin{abstract}
Earth's atmosphere is in a state far from thermodynamic
equilibrium. For example, the large scale equator-to-pole temperature
gradient is maintained by tropical heating, polar cooling, and a
midlatitude meridional eddy heat flux predominantly driven by
baroclinically unstable weather systems. Based on basic thermodynamic
principles, it can be shown that the meridional heat flux, in combination
with the meridional temperature gradient, acts to
maximize entropy production of the atmosphere. In fact, maximum
entropy production (MEP) has been successfully used to explain the
observed mean state of the atmosphere and other components of the
climate system. However, one important feature of the large scale
atmospheric circulation is its often non-Gaussian variability about
the mean. This paper presents theoretical and observational evidence
that some processes in the midlatitude atmosphere are significantly
non-Gaussian to maximize entropy production. First, after introducing
the basic theory, it is shown that the skewness of sea surface winds,
damped by nonlinear surface drag, are consistent with the MEP
principle.  Then it is pointed out that the observed $k^{-3}$
wavenumber spectrum of long planetary waves in the midlatitudes can be
roughly explained by maximizing the meridional eddy heat flux (and
related entropy production) of unstable baroclinic Eady waves.
Finally observational evidence is presented that the meridional eddy
heat flux is increased by non-Gaussian variability in meridional wind
and temperatures anomalies.
\end{abstract}

\newpage

\pagestyle{plain}


\section{Introduction}
The atmospheric system is a high-dimensional and highly complex
dynamical system with very many nonlinearly interacting modes. 
One important feature of the large scale (predominantly low-frequency)
atmospheric circulation is its non-Gaussianity. Understanding
non-Gaussianity has become an important objective in weather/climate
research, because weather and climate risk assessment depends on
knowing and understanding the tails of probability distributions. In
particular, there is broad consensus that the most hazardous effects
of climate change are related to a potential increase (in frequency
and/or intensity) of extreme weather and climate events populating the
tails of often non-Gaussian distributions.

Various mechanisms behind the observed atmospheric non-Gaussian
statistics have been proposed. The mechanisms are multifaceted (e.g.,
nonlinear dynamics, multiplicative noise, cross-frequency coupling)
and scattered in the literature. \cite{Sura:2014} provide a
comprehensive review, focusing on atmospheric synoptic and
low-frequency variability. However, while there are many mechanisms
contributing to non-Gaussian atmospheric statistics, there does not
exist a deeper grand principle beyond the diversity of phenomena. Yet
it is not unreasonable to believe that there may exist a deeper grand
principle beyond the diversity of phenomena. For example, one
possibility is that atmospheric and/or climate statistics are
collectively non-Gaussian to maximize a specific functional of the
weather/climate system. One possible candidate to explore would be the
entropy production \cite[e.g.,][]{Paltridge:1975,Ozawa:2001,
Ozawa:2003,Kleidon:2008}. That is, the climate system may exhibit a
specific collective non-Gaussianity to maximize the overall entropy
production.

Maximum entropy production (MEP) has already been successfully (and
extensively) used to explain the observed {\em mean} state of the
atmosphere and climate, such as the meridional distribution of mean
air temperature, cloud cover, and meridional heat transport
\cite[][]{Paltridge:1975}. While the initial idea goes back to
\cite{Paltridge:1975}, excellent summaries/reviews have been provided
by many authors \cite[e.g.,][]{Ozawa:2001,Ozawa:2003,Kleidon:2008}.
At this point it is important to realize that, despite many attempts,
MEP has (so far) no generally accepted theoretical basis. That is,
while it makes physical sense that a system {\em not} in its unique
thermodynamic equilibrium state wants to get there as fast as possible
by maximizing entropy production, this physically reasonable statement
does not constitute a universal proof that the MEP principle is
generally valid in nature. A better term would be MEP conjecture, but
for the sake of simplicity we use the words ``principle'' and
``conjecture'' interchangeably.

In this paper we propose that MEP may also point a way forward towards
a unified perspective of {\em non-Gaussianity} in weather and climate
variability. The basic theory (equations governing the time rate of
change of entropy of an open fluid system) is derived and explained in
section \ref{theory}. In contrast to previous work the theory is
formulated to highlight the potential impact of probability
distributions on entropy production. In section \ref{applications} we
use MEP to study a simplified conceptual model of the midlatitude
atmospheric circulation. The model consists of a midlatitude volume
above the boundary layer. The baroclinically unstable atmosphere
within that volume, which is frictionally damped by the surface drag
due to turbulence in the atmospheric boundary layer, transports heat
poleward to keep the average temperatures in the tropics and in the
polar zone constant. As a first result we will see that the observed
skewness of sea surface winds, damped by nonlinear surface drag, is
consistent with MEP. While it is already known that the product
$\overline{F}\Delta{T}$ of the meridional heat flux $\overline{F}$ and
the meridional temperature gradient $\Delta{T}$ has to be maximized to
be in accordance with the MEP principle\footnote{Note that if the
meridional temperature gradient $\Delta{T}$ is assumed to be constant,
it is the meridional heat flux $\overline{F}$ alone that has be maximized.},
a novel finding is that the observed $k^{-3}$ wavenumber spectrum
of long waves can be roughly explained by maximizing the meridional
eddy heat flux $\overline{v'T'}$ of unstable baroclinic Eady
waves. Next it is shown that the meridional eddy heat flux is
increased by non-Gaussian variability in meridional wind and
temperatures anomalies $v'$ and $T'$.  Overall that means, as
discussed in the final section \ref{conclusions}, that MEP is
consistent with the observed non-Gaussian variability in the
midlatitude atmosphere, and may point a way forward towards a unified
perspective of non-Gaussianity in weather and climate variability.

\section{Theory}
\label{theory}
\subsection{Basics}
\vspace{-3ex}

In the following let us derive and discuss the basic equations
governing the time rate of change of entropy of an {\em open} fluid system
[in the style of \cite{Ozawa:2001}]. Here {\em open} means that the system
exchanges heat and momentum, but not overall mass, with its
surrounding. The starting equation is the time derivative
\begin{equation}
\dot{S}_{sys}
= \frac{d}{dt}\left[\int \rho s dV \right]
= \int \frac{\partial{(\rho s)}}{\partial{t}}  dV + \int \rho s  v dA  \,\,,
\label{entropy-1}
\end{equation}
where $\rho$ is the density of the fluid, $s$ is the entropy per unit
mass, $v$ is the normal (positive outward) component of the fluid
velocity at the surface, $V$ is the volume of the system, and $A$ is
the surface bounding the system. It is important to note that the
boundary flux term $\int{\rho s v}dA$ is needed to maintain the
overall (i.e., integrated) mass of the otherwise open fluid system.
In the discussion of entropy production by \cite{LandauLifshitz:1966}
only the volume integral appears because they consider a closed
system. That is, in \cite{LandauLifshitz:1966} the mass of the systems
is automatically conserved because there is no exchange with its
surrounding at all (that is, $v$ in the boundary flux term is zero by
construction).

Next, using the continuity equation $\partial{\rho}/\partial{t}=-\nabla
\cdot(\rho \mathbf{v})$ we obtain
\begin{equation}
\frac{\partial{(\rho s})}{\partial{t}}
=\rho \frac{\partial{s}}{\partial{t}} + s \frac{\partial{\rho}}{\partial{t}}
= \rho \frac{\partial{s}}{\partial{t}} - \nabla \cdot (\rho s \mathbf{v})
+ \rho \mathbf{v} \cdot \nabla s \,\,,
\end{equation}
which we can use, together witch Gauss' theorem, to rewrite
Eq.~(\ref{entropy-1}) to get
\begin{equation}
\dot{S}_{sys}
= \int \rho \left[\frac{\partial{s}}{\partial{t}} 
+ \mathbf{v} \cdot \nabla s   \right]dV
= \int \rho \frac{ds}{dt} dV \,\, .  
\label{entropy-2}
\end{equation}
Note that the expression in square brackets is just $ds/dt$, the total
derivative of the entropy per unit mass following the fluid motion.
However, we know that $ds=dq/T$, where $dq$ is the heat flux (per unit
mass) into the small volume element $dV$ and $T$ is its temperature.
In addition, from the first law of thermodynamics we know that 
$dq=du + pd(1/\rho)$ (here $u$ is the internal energy per unit mass and $p$
is pressure). Therefore,
\begin{equation}
\frac{ds}{dt} 
= \frac{1}{T} \left( \frac{du}{dt} +  p \frac{d(1/\rho)}{dt} \right) \,\, .
\end{equation}
Using this in Eq.~(\ref{entropy-2}), the fact that 
$du/dt=\partial{u}/\partial{t} + \mathbf{v} \cdot \nabla u$, and the continuity
equation in the form $d(1/\rho)/dt= (1/\rho) \nabla \cdot \mathbf{v}$ , we obtain
\begin{equation}
\dot{S}_{sys} 
= \int \frac{1}{T} \left[ \rho \frac{\partial{u}}{\partial{t}}
+ \rho \mathbf{v} \cdot \nabla u + p \nabla \cdot \mathbf{v} \right] dV \,\, .
\end{equation}
by using
\begin{equation}
\rho \frac{\partial{u}}{\partial{t}} + \rho \mathbf{v} \cdot \nabla u 
= \frac{\partial{(\rho u)}}{\partial{t}} + \nabla \cdot (\rho u \mathbf{v})  
\end{equation}
and $u=cT$ (where $c$ is the specific heat at constant volume)
we finally get
\begin{equation}
\dot{S}_{sys} 
= \int \frac{1}{T} \left[\frac{\partial{(\rho c T)}}{\partial{t}}
+ \nabla \cdot (\rho c T \mathbf{v})  + p \nabla \cdot \mathbf{v} \right] dV
\equiv \int \frac{Q}{T} dV \,\, .
\label{entropy-3}
\end{equation}
Equation (\ref{entropy-3}) describes the rate of entropy change of an
open fluid system in terms of pressure, temperature, density, and velocity fields.
The term in square brackets is the local {\em diabatic} heating rate per unit
volume $Q$ which can be rewritten in terms of the convergence of a diabatic
heat flux density $\mathbf{F}$ due to fluid motion (turbulence) and the 
heating rate (per unit volume) $\Phi$ representing the rate of
dissipation of kinetic energy by viscosity:
\begin{equation}
Q= - \nabla \cdot \mathbf{F} + \Phi \,\, . 
\label{entropy-4}
\end{equation}
The heat flux $\mathbf{F}$ includes all diabatic heat transport
processes associated with turbulent fluid motion (i.e., heat
conduction, latent heat transport).  It is important to note that
$\mathbf{F}$ does {\em not} include the advective heat flux because
advection (i.e. coherent motion of fluid) is intrinsically a
reversible process (one can reverse the heat transport by reversing
the movement). We could include a reversible advective heat flux in
$\mathbf{F}$ but it would results in {\em no} contribution to entropy
production of the whole system: moving the internal energy (i.e. heat)
of fluid from one place to another alone will not change the system's
overall entropy. However, the advection of, for example, hot fluid
into a colder region results in entropy production by heat conduction
across the frontal regions with strong temperature gradients (as we
will see from the overall entropy production equation next).

While Eq.~(\ref{entropy-3}) describes the entropy production of the
open fluid system, the entropy of the surrounding system will also
increase by the heat flux $\mathbf{F}$ from the fluid system through the boundary.
The rate of entropy change of the surrounding system $\dot{S}_{surr}$
is given by the surface integral
\begin{equation}
\dot{S}_{surr} = \int \frac{F}{T} dA \,\,\ ,
\label{entropy-5}
\end{equation}
where $F$ is the normal component (defined as positive outward) of the
heat flux $\mathbf{F}$. Now we are in the position to formulate the
budget for entropy production (due to turbulence in the fluid system) of the 
whole system as the sum of Eqs.~(\ref{entropy-3}) and (\ref{entropy-5}):
\begin{equation}
\dot{S}_{turb}
= \int \frac{- \nabla \cdot \mathbf{F} + \Phi}{T} dV
+ \int \frac{F}{T} dA \,\, .
\label{entropy-6}
\end{equation}
Using Gauss' theorem we can rewrite the surface integral as
\begin{equation}
\int \frac{F}{T} dA = \int \nabla \cdot \left( \frac{\mathbf{F}}{T} \right) dV
= \int \frac{\nabla \cdot \mathbf{F}}{T} dV 
+ \int \mathbf{F} \cdot \nabla \left( \frac{1}{T} \right) dV \,\, ,
\end{equation}
to obtain our final entropy production equation for the whole system 
\begin{equation}
\dot{S}_{turb}=\int \mathbf{F}\cdot\nabla\left( \frac{1}{T} \right) dV 
+ \int \frac{\Phi}{T} dV  \,\, .
\label{entropy-7}
\end{equation}
Equation (\ref{entropy-7}) describes the rate of entropy change of
the whole system due to motion/turbulence in the fluid system.
The first term is the rate of entropy increase by heat conduction (i.e.
thermal dissipation), and the second is that by viscous dissipation.

Finally, let us look into a steady state system in some more detail. Then
the entropy of the fluid system remains constant and Eq.~(\ref{entropy-6})
becomes
\begin{equation}
\dot{S}_{turb,st}
= \int \frac{F}{T} dA \,\, ,
\label{entropy-8}
\end{equation}
where the subscript $st$ denotes that the fluid systems is in a steady state.
This equation tells us that in a steady state the entropy produced by irreversible
processes (i.e., thermal and viscous dissipation) in a turbulent fluid
is entirely discharged to the surrounding system through a boundary heat flux $F$.   
That is, in a steady state we have
\begin{equation}
\dot{S}_{turb,st}=\int \mathbf{F}\cdot\nabla\left( \frac{1}{T} \right) dV 
+ \int \frac{\Phi}{T} dV  
= \int \frac{F}{T} dA = \textrm{Maximum} \,\, ,
\label{entropy-9}
\end{equation}
where we ultimately also {\em assumed} that the fluid system maximizes
the entropy production. It has been already mentioned in the
introduction that, while it makes physical sense that a system not in
its unique thermodynamic equilibrium state wants to get there as fast
as possible by maximizing entropy production, this statement does not
constitute a universal proof that the MEP principle is generally
valid. However, it is possible to show that the global distribution of
key climate variables and several maximum transport properties
suggested in the literature are consistent with Eq.~(\ref{entropy-9})
\cite[][]{Paltridge:1975,Ozawa:2001,Ozawa:2003}, making MEP a
physically reasonable tool for climate research.

\subsection{Maximum entropy production: probability distributions}
\vspace{-3ex}

It should be noted that Eq.~(\ref{entropy-9}) describes the
instantaneous state of the fluid system. However, describing a
turbulent fluid by its instantaneous state does not make much
statistical sense. That is, we should interpret Eq.~(\ref{entropy-9})
in a time averaged framework:
\begin{equation} 
\overline{\dot{S}}_{turb,st}
=\int \overline{ \mathbf{F}\cdot\nabla\left( \frac{1}{T} \right) } dV 
+ \int \overline{ \left( \frac{\Phi}{T} \right) } dV  
= \int \overline{ \left( \frac{F}{T} \right) } dA = \textrm{Maximum} \,\, ,
\label{entropy-10}
\end{equation}
where the overbar denotes a time mean. Remember that the time mean
$\overline{x}$ of the variable $x$ can also be written in terms of its
probability density function (PDF) $p(x)$: $\overline{x}= \int x p(x)
dx$.  That means that we can rewrite Eq.~(\ref{entropy-10}) in terms
of the PDFs of the entropy increases due to thermal dissipation,
viscous dissipation, and boundary heat fluxes. If we define the new
variables $X_1 \equiv \mathbf{F}\cdot\nabla\left( 1/T \right)$,
$X_2\equiv \Phi/T $, and $X_3 \equiv F/T$ we can rewrite the MEP
proposition as
\begin{eqnarray}
\overline{\dot{S}}_{turb,st}
&=& \int \left( \int X_1 p(X_1) dX_1 \right) dV
+ \int \left( \int X_2 p(X_2) dX_2 \right) dV \nonumber \\ 
&=& \int \left( \int X_3 p(X_3) dX_3 \right) dA \nonumber \\
&=& \textrm{Maximum} \,\, .
\label{entropy-11}
\end{eqnarray}
In this equation it is apparent that the entropy production depends on
the PDFs of thermal dissipation $X_1$, viscous dissipation $X_2$, and
boundary heat fluxes $X_3$. Let us now apply the MEP principle
(\ref{entropy-11}) to explain some of the non-Gaussian features
observed in the midlatitude atmosphere.

\section{Applications}
\label{applications}
In the following discussion we consider a simplified conceptual model
of the atmospheric circulation depicted in Fig.~\ref{Figure-1}. The
atmosphere is composed of three regions: a tropical zone, the
midlatitudes, and a polar zone. The average temperature in the tropics
is $T_{Trop}$ and the average temperature in the the polar zone is
$T_{Pole}$. In the tropics there is a net input of radiation/heat,
whereas in polar regions there is a net output. To keep the average
temperatures in the tropics and in the polar zone constant the
midlatitude atmosphere transports heat poleward through its
(predominantly horizontal) circulation, characterized through zonal
and meridional velocities $u$ and $v$, and temperature $T$.  The
poleward heat flux consists of a flux from the tropics into the
midlatitudes $F_{TM}$, and a flux from the midlatitudes into the polar
zone $F_{MP}$. Below the free midlatitude atmosphere we have a
turbulent boundary layer that exerts a drag on the surface winds and
induces a dissipative heating rate $\Phi$.

In the following we first consider the entropy production through
viscous dissipation in the boundary layer. Then we study entropy
production in a midlatitude volume (the shaded region) above the
boundary layer.

\subsection{Viscous dissipation}
\vspace{-3ex}

\begin{figure}[H]
\begin{center}
\includegraphics[scale=1.0,angle=0]{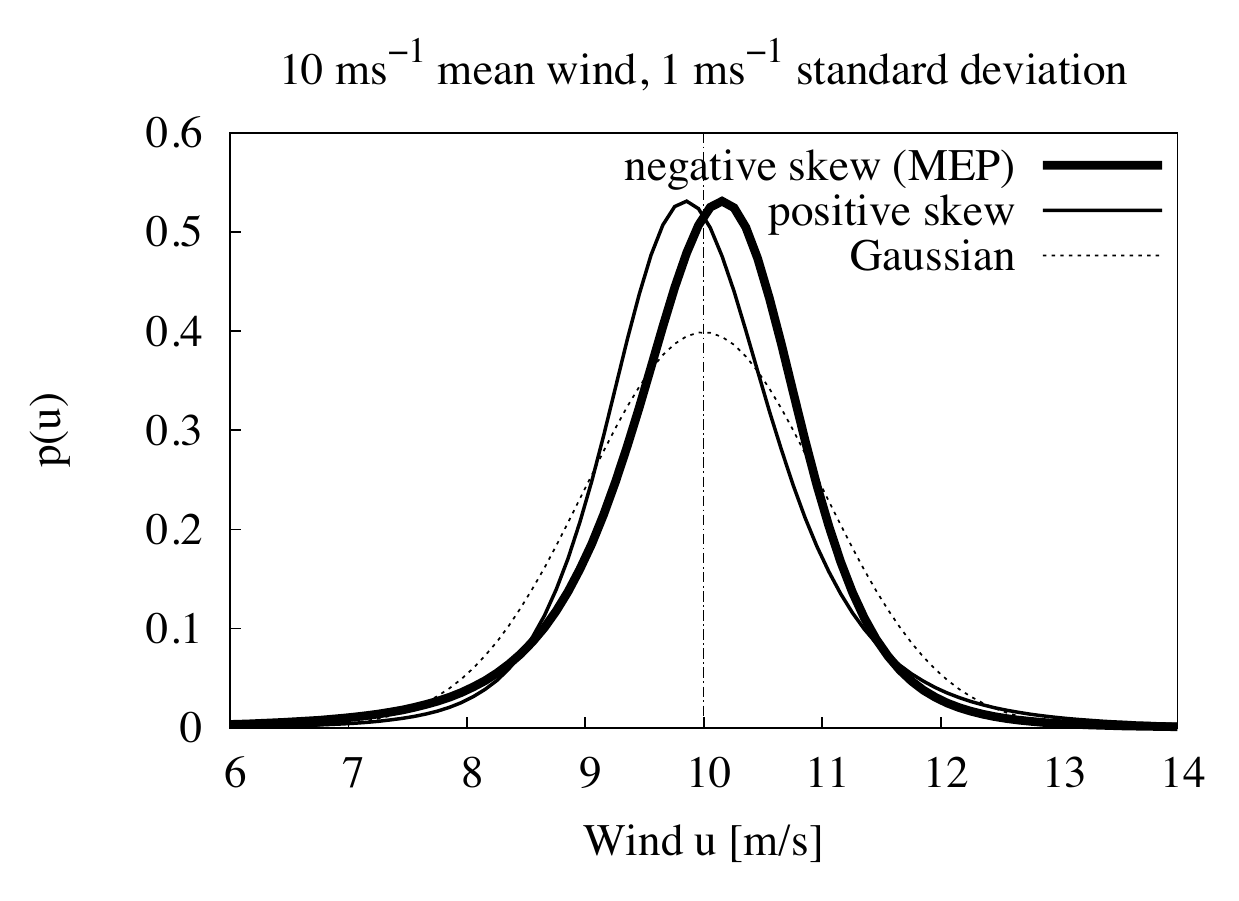}
\end{center}
\caption{\label{CAM-pdf}
PDFs generated by a univariate linear stochastic differential equation
with correlated additive and linear multiplicative (CAM) noise
\cite[][]{Sardeshmukh:2009,Sura:2011,Sura:2013}.
The thick solid line denotes a PDF with negative skewness, the thin solid line 
is a PDF with positive skewness, and the dashed line represents a Gaussian; 
all three PDFs have the same 10 ms$^{-1}$ mean and unit variance.
From the PDFs we can calculate the entropy production $\propto\overline{u^2 |u|}$.
The PDF with negative skewness maximizes $\overline{u^2|u|}$.}
\end{figure}

As a simple example let us consider a well-mixed atmospheric boundary
layer where the wind speeds (and temperature) are approximately
independent of height.  According to Monin-Obukhov similarity theory,
the zonal and meridional components of the surface stress vector are
$\tau_x=\rho c_D u \sqrt{u^2+v^2}$ and $\tau_y= \rho c_D v
\sqrt{u^2+v^2}$, where $u$ and $v$ are the zonal and meridional
surface wind components, $c_D$ is the (nondimensional) drag
coefficient, and $\rho$ is the density of air.  Then the rate of
velocity changes (now in vector notation) due to viscous dissipation
are
\begin{equation}
\frac{\partial{\mathbf{u}}}{\partial{t}} = \dots 
- \frac{c_D}{h}\mathbf{u}|\mathbf{u}|\,\,\, ,
\end{equation}
where $h$ is the depth of the boundary layer (the dots symbolize all the
other terms in the momentum budget). As the energy lost to
viscosity is ultimately converted to heat, the (instantaneous) viscous
heating rate per unit volume in the boundary layer is
\begin{equation}
\Phi= \frac{\rho c_D}{h} \mathbf{u}^2 |\mathbf{u}| \,\, .
\end{equation}
Then the averaged MEP proposition (\ref{entropy-10},
\ref{entropy-11}) simplifies to
\begin{equation}
\overline{\dot{S}}_{turb,st}
= \int \overline{ \left( \frac{\Phi}{T} \right) } dV 
\propto \int \overline{ \mathbf{u}^2 |\mathbf{u}| } dV 
= \textrm{Maximum} \,\, .
\end{equation}
That is, locally surface winds have to maximize $\overline{\mathbf{u}^2 |\mathbf{u}|}$. 
If, for the sake of conceptual simplicity, we also assume a monodirectional
wind $u$, we obtain 
\begin{equation}
\overline{\mathbf{u}^2 |\mathbf{u}|} 
= \overline{u^2 |u|}
= \int u^2 |u| p(u) du
= \textrm{Maximum} \,\, .
\end{equation}
To study this equation we use a general parametric PDF $p(u)$ that has
been shown to describe non-Gaussian atmospheric variability very well
\cite[][]{Sardeshmukh:2009,Sura:2011,Sura:2013}. The non-Gaussian PDF
is generated by a univariate linear stochastic differential equation
with correlated additive and linear multiplicative (CAM) noise. In
particular, it is the CAM noise that is responsible for skewness and
kurtosis of the related PDF. As an example, we use a PDF with a 10
ms${^{-1}}$ mean wind and 1 ms${^{-1}}$ standard deviation.  The
non-Gaussianity can be controlled by changing the CAM noise parameter,
while keeping mean and standard deviation constant. If the additive
and multiplicative noise are positively correlated, the skewness is
positive; the reverse (negative skewness) is true for negatively
correlated additive and multiplicative noise. If we don't have
multiplicative noise, the PDF is Gaussian. Representative plots are
shown in Fig.~\ref{CAM-pdf}. The thick solid line denotes a PDF with
negative skewness, the thin solid line is a PDF with positive
skewness, and the dashed line represents a Gaussian; all three PDFs
have the same mean and variance. Given the PDFs we can calculate the
entropy production $\propto\overline{u^2 |u|}$. It turns out that, in
general, the PDF with negative skewness maximizes $\overline{u^2|u|}$. 
That is, for positive mean winds MEP predicts negative skewness. For
negative mean winds MEP favors positive skewness (not shown). Sure
enough, this link between mean and skewness is consistent with
observations.

\begin{figure}[H]
\begin{center}
\includegraphics[scale=1.1,angle=0]{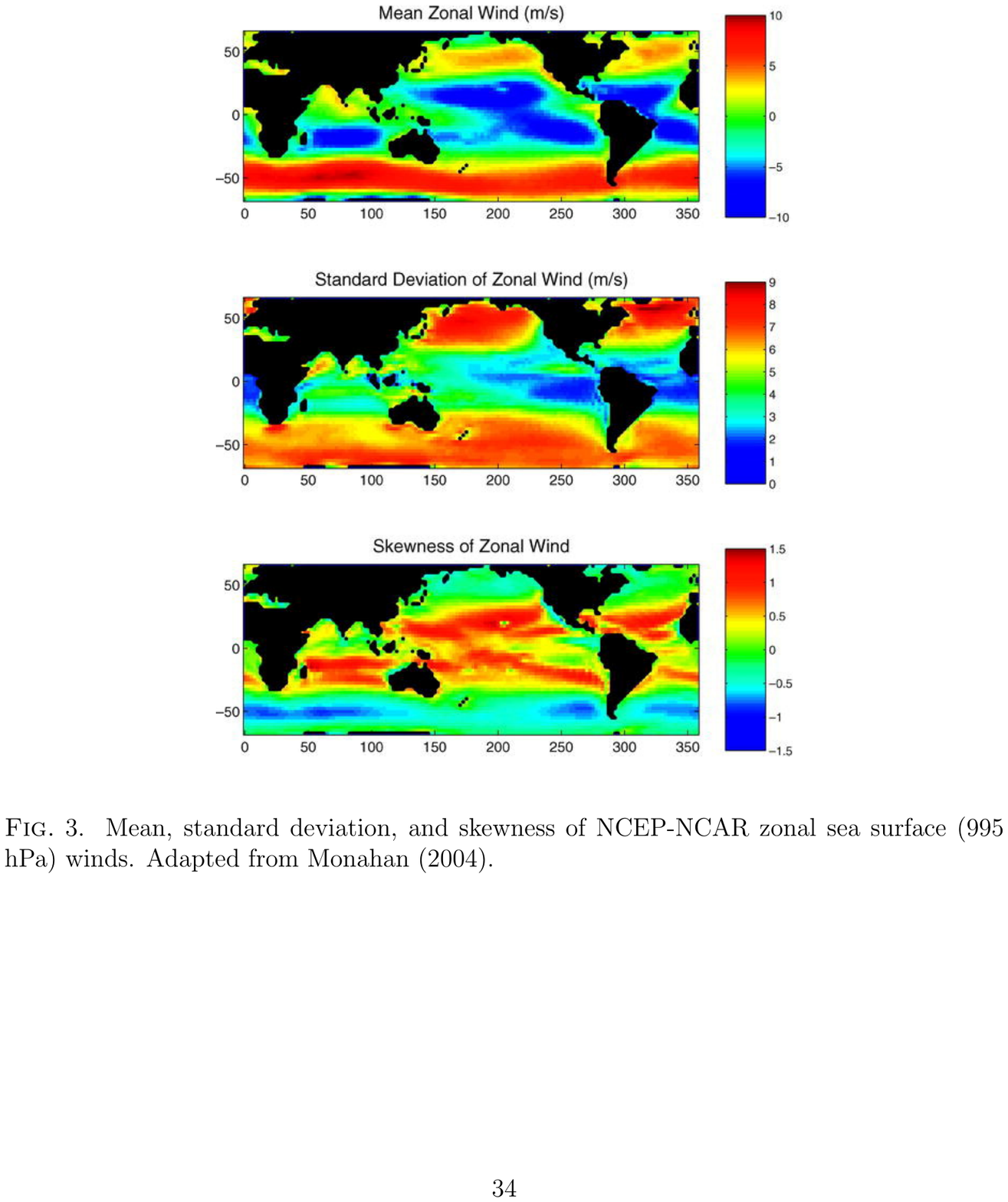}
\end{center}
\caption{\label{Monahan2004} Mean, standard
deviation, and skewness of NCEP-NCAR zonal sea surface (995 hPa)
winds. Adapted from \cite{Monahan:2004b}.}
\end{figure}

For example, Fig.~\ref{Monahan2004} shows the mean, standard
deviation, and skewness of NCEP-NCAR zonal sea surface (995 hPa)
winds. The mean sea surface zonal wind field displays the familiar
tropical easterlies and midlatitude westerlies. The standard deviation
shows variability minima in the tropics/subtropics and maxima in the
midlatitude storm tracks. The skewness of the zonal wind is generally
positive in the tropics and negative in the midlatitudes.  In fact,
there exists a linear relationship between the mean and skewness of
the zonal winds: positive means winds are related to negative skewness
and vice versa. The same relationship holds for meridional winds; see
\cite{Monahan:2004b} for more details.

While here we derived the basic link between mean and skewness of sea
surface winds using the MEP proposition, the dynamical principle
behind the observed anticorrelation can also be understood rather
easily \cite[][]{Monahan:2004b}. As with MEP, the main building block
is the fact that the stress in the boundary layer is nonlinear. Now
the basic understanding of the link between the mean and the skewness
of the sea surface wind components is straightforward. First suppose
that the mean wind is zero and that the fluctuations are equally
likely to be positive and negative. Then both the mean and skewness of
the resulting wind will be zero. Now suppose that the mean wind is
positive. Because of the nonlinear surface drag, positive wind
anomalies will experience stronger friction than negative ones.
Therefore, the winds will be negatively skewed. The opposite is true
if we consider a negative mean wind. That is, the non-Gaussianity of
the sea surface winds is due to the nonlinear boundary layer drag.
Over land the situation is a bit more complicated because the boundary
layer is influenced more strongly by the diurnal cycle than over sea
\cite[][]{He:2010,Monahan:2011}.

\subsection{Meridional heat transport}
\vspace{-3ex}

As the next example let us now neglect the effect of the boundary
layer and focus on the horizontal heat fluxes (Fig.~\ref{Figure-1}).
In our conceptual model of the atmosphere that means that the flux
from the tropics into the midlatitudes equals the flux from the
midlatitudes into the polar zone: $F_{TM}=F_{MP} \equiv F$. Then the
MEP proposition (\ref{entropy-10}, \ref{entropy-11}) becomes
\begin{equation}
\overline{\dot{S}}_{turb,st}
=\frac{\overline{F}}{T_{Pole}} - \frac{\overline{F}}{T_{Trop}}
=\frac{\Delta{T}\overline{F}}{T_{Pole}T_{Trop}} 
= \textrm{Maximum} \,\,\ ,
\label{entropy-circulation}
\end{equation}
where $\Delta{T}=T_{Trop} - T_{Pole}$. That is, the product of the
mean meridional heat flux $\overline{F}$ and the meridional
temperature gradient $\Delta{T}$ has to be maximized to be in
accordance with the MEP principle. If we treat the temperatures in the
zones/reservoirs as constant $\Delta{T}$ is also constant, and the
mean meridional heat flux $\overline{F}$ alone has to be maximized for
the global atmosphere to be in a state of maximum entropy production.
However, it is important to note that the tropics to pole temperature
difference $\Delta{T}$ is actually a function of the mean heat flux
$\overline{F}$: $\Delta{T}=\Delta{T}(\overline{F})$
\cite[][]{Ozawa:2001,Ozawa:2003}.  To understand this, consider the
extreme case of a motionless atmosphere.  In this static state the
meridional heat flux vanishes: $\overline{F} \approx 0$.  Then the
tropics will heat up and the poles will cool down, resulting in an
increased (decreased) thermal emission from the tropics (poles),
respectively, to close the energy imbalance in each region.
Therefore, the temperature difference $\Delta{T}$ will be largest in
the static state. If we now allow $\overline{F}$ to increase from
zero, $\Delta{T}$ will decrease. On the other hand, if we have extreme
atmospheric mixing resulting in a very large $\overline{F}$, the
temperature difference $\Delta{T}$ will be very small. Overall this
means that $\Delta{T}(\overline{F})$ is a monotonic decreasing function of
$\overline{F}$ \cite[][]{Ozawa:2001,Ozawa:2003}.  Now, because the
entropy production $\overline{\dot{S}}_{turb,st}$ in
(\ref{entropy-circulation}) is proportional to the product
$\Delta{T}(\overline{F})\overline{F}$, it must have a maximum between
the two extreme cases $\overline{F}=0$ (static state with no mixing)
and $\Delta{T}(\overline{F})=0$ (extreme mixing). According to the MEP
principle (\ref{entropy-10}, \ref{entropy-11}), this maximum
corresponds to the optimal mean state of the atmospheric circulation.
Starting with \cite{Paltridge:1975} it has been shown in many papers
[see \cite{Ozawa:2001,Ozawa:2003} and \cite{Kleidon:2008} for
excellent reviews] that the observed mean state of the climate system
is indeed maximizing entropy production.

As mentioned earlier, in its original formulation (\ref{entropy-9},
\ref{entropy-10}) $\overline{F}$ does not include advective heat
fluxes because advection is a reversible process.  However, in the
case of our box model we have to take a closer look. The midlatitude
atmosphere is predominantly transporting heat by turbulent advection
$\overline{v'T'}$ induced by baroclinic instability. As baroclinic
instability is effectively just replacing cold high-density air from
high altitudes (and latitudes) with warm low-density air from low
altitudes (and latitudes), it is obvious that there is no entropy
produced in that process. If the turbulent (eddy) heat flux does not
produce entropy, where is the entropy produced in our box model?  In a
similar model for small scale thermal convection (like in a lava lamp)
the MEP equation is identical to (\ref{entropy-circulation}), where
the temperatures are then related to the heating at the bottom and
cooling at the top \cite[][]{Ozawa:2001,Ozawa:2003}. Within the bulk
of the fluid the convection (i.e., heat advection) does not produce
entropy. Only in the thin thermal boundary layers the temperature
gradients are, on average, large enough to establish an irreversible
heat conduction flux $\overline{F}$ out of the fluid volume that
actually contributes to the entropy production. In our atmospheric
model (Fig.~\ref{Figure-1}) we do not have thermal boundary layers
between the boxes, so the close analogy breaks down here. However,
because we have a mass flux between the boxes due to the atmospheric
general circulation (keeping the net masses of our boxes constant,
though), we implicitly assumed that tropical and polar boxes are well
mixed to have constant temperatures. That is, the box model setup
implies that entropy is produced mainly through mixing, and that the
entropy production through mixing equals the rate predicted by
(\ref{entropy-circulation}) with advection included in $\overline{F}$
[as in the multiple box model used \cite{Paltridge:1975}].

\subsubsection{Meridional heat transport through baroclinic instabilty}
\vspace{-3ex}

In the midlatitude atmosphere most of the meridional heat transport is
done by unstable baroclinic eddies (i.e., weather systems). Therefore,
the meridional heat flux is dominated by the eddy covariance of
temperature anomalies $T'$ and meridional velocity anomalies $v'$:
$\overline{F} \approx c\overline{v'T'}$, where $c$ is the specific
heat at constant volume. While we know that the atmospheric
circulation maximizes entropy production and therefore, if we assume
for the sake of simplicity that the meridional temperature gradient
$\Delta{T}$ is constant, the heat flux $\overline{F}$, we do not know
much about the dynamics behind the maximization. To do that, we need
to have a dynamical model representing baroclinic instability in our a
midlatitude volume (the shaded region above the boundary layer) in
Fig.~\ref{Figure-1}.

The simplest model of baroclinic instability, which represents the
essential instability process in its purest and mathematically
feasible form, was introduced by \cite{Eady:1949}. The details of
Eady's model can be found in all standard textbooks on dynamical
meteorology and geophysical fluid dynamics
\cite[e.g.,][]{Holton:2004,Pedlosky:1987,Gill:1982,Vallis:2006}. That
is, in the following discussion we will use Eady's model to represent
unstable baroclinic waves and the related poleward heat flux in our
midlatitude volume. In the previous section we have seen that the
tropics to pole temperature gradient is actually a function of the
meridional heat flux: $\Delta{T}=\Delta{T}(\overline{F})$. However, in
the following discussion of the heat flux induced by unstable Eady
waves we consider the temperature gradient as constant. That means
that we are strictly speaking not maximizing entropy production but
the heat flux. In this simplified setting maximizing entropy
production and maximizing heat flux are equivalent. Nevertheless, in
the real atmosphere these two principles are {\em not} identical any
more because $\Delta{T}=\Delta{T}(\overline{F})$. Thus in the
following treatment of the Eady model we consider a maximum flux
principle that we derived as a special (i.e., simplified)
case of the MEP principle, keeping in mind that they are not
equivalent. 

\begin{figure}[H]
\begin{center}
\includegraphics[scale=1.0,angle=0]{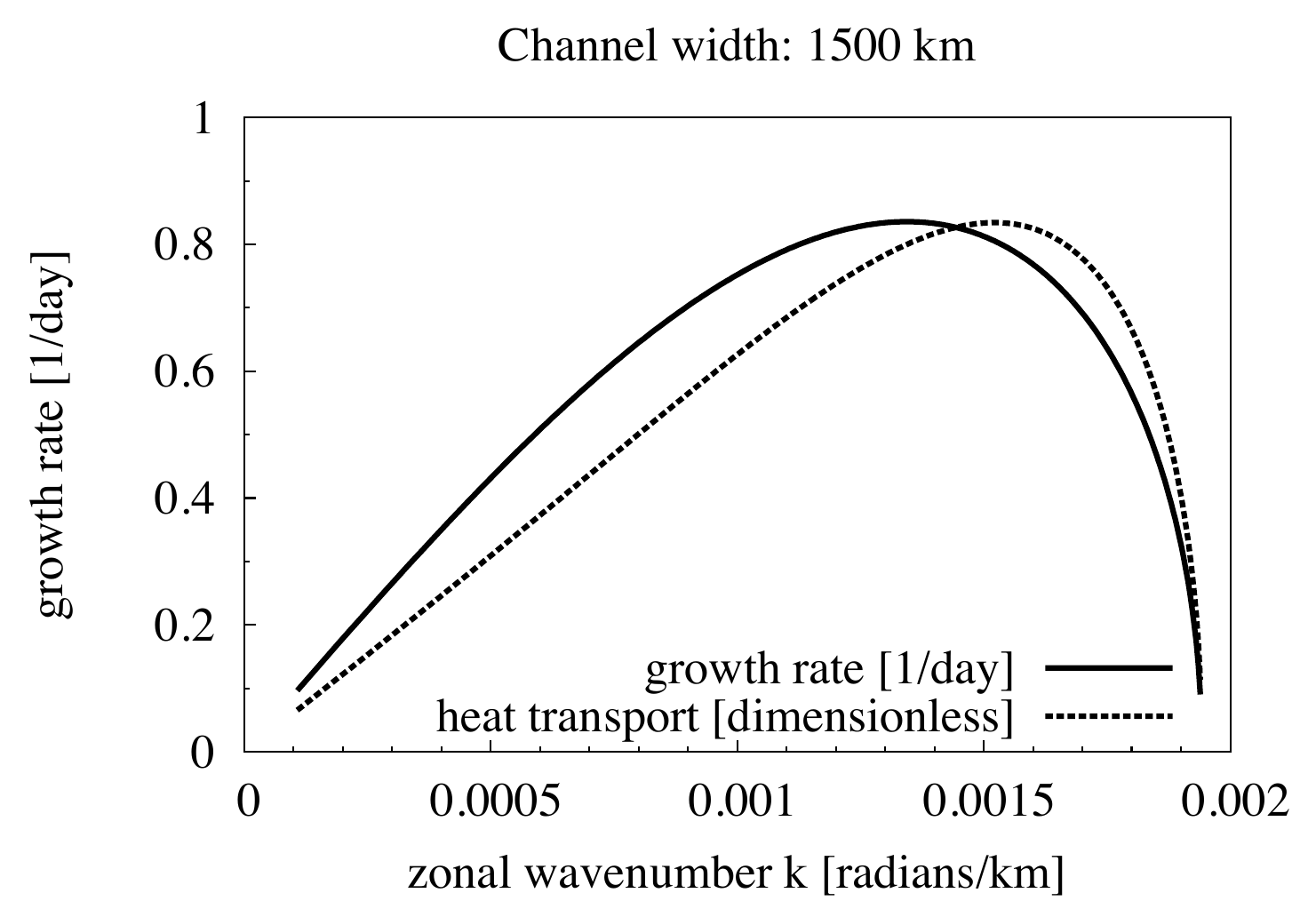}
\end{center}
\caption{\label{Eady1} Growth rate $c_ik$ (solid line) and dimensionless 
eddy heat flux $\overline{v'T'} \propto c_i \alpha^2 k$ (dotted line) of 
unstable Eady wave as a function of zonal wavenumber $k$ for a meridional 
channel width of 1500 km. Note that the eddy heat flux is  arbitrarily scaled 
to have the same maximum value as the growth rate to make the shape of the 
graphs easily comparable.}
\end{figure}

In a nutshell, the Eady model is a quasi-geostrophic and continuously
stratified baroclinic atmosphere in a $f$-plane channel of width
$L_y$. The basic state density is constant, the vertical shear of the
basic state zonal flow is constant
($\partial{\overline{u}}/\partial{z} = \Lambda = \textrm{constant}$),
and it has rigid lids at the bottom ($z=0$) and the top ($z=H$) of the
atmosphere. Then the linearized potential vorticity equation and the
first law of thermodynamics in terms of the streamfunction
$\psi=\overline{\psi}+\psi'$ [i.e., $\partial{\psi}/\partial{y}=-u$
($u$ is zonal wind), $\partial{\psi}/\partial{x}=v$ ($v$ is meridional
wind), $\partial{\psi}/\partial{z}= RT/f_0H$ ($T$ is temperature), and
$R$ = 287 J K$^{-1}$ kg$^{-1}$ is the gas constant for dry air] are 
\begin{equation}
\left(\frac{\partial}{\partial{t}}
+ \overline{u}\frac{\partial}{\partial{x}} \right)
\left(\nabla^2\psi' + \epsilon \frac{\partial^2{\psi'}}{\partial{z}^2}
\right)=0
\end{equation}
and
\begin{equation}
\left(\frac{\partial}{\partial{t}} 
+ \overline{u}\frac{\partial}{\partial{x}} \right)
\frac{\partial{\psi'}}{\partial{z}}
- \frac{\partial{\psi'}}{\partial{x}}\frac{\partial{\overline{u}}}{\partial{z}}
=0 \,\, ,
\end{equation}
where $\epsilon=f_0^2/N^2$ ($f_0$ is the Coriolis parameter and $N$ is
the buoyancy frequency). Solutions can be found in the form
$\psi'(x,y,z,t)=\Psi(z)\cos(ly)\exp[ik(x-ct)]$, where $\Psi(z)$ is a
complex amplitude and $c=c_r + ic_i$ is a complex phase speed.
Obviously, the flow is unstable with a growth rate $kc_i$ if $c$ has
an imaginary part $c_i > 0$. The stability criterion can be obtained
from the dispersion relation $c=\Lambda H /2 \pm \Lambda H /2 \left[1
  - 4\coth(\alpha H)/(\alpha H) + 4/(\alpha H)^2 \right]^{1/2}$, where
$\alpha^2=(k^2+l^2)/\epsilon$: The flow is unstable for $\alpha <
\alpha_c$, where the critical value $\alpha_c$ is given by $\alpha_c
H/2=\coth(\alpha_c H/2)$ or $\alpha_c \approx 2.4/H$.  For typical
atmospheric values ($H$ = 7 km, $N$ = 1.2$\times 10^{-2}$ s$^{-1}$,
$f_0$ = $10^{-4}$ s$^{-1}$, $\Lambda$ = 8$\times 10^{-3}$ s$^{-1}$) we
obtain $\alpha_c$ = 0.343 km$^{-1}$.  If we now calculate the zonal
wavenumber $k$ of maximum growth rate for a channel width $L_y=$ 1500
km and allow only half a wavelength in the meridional direction (i.e.,
$l=\pi/L_y$), we obtain $\lambda_g \approx 4500$ km (or
$k_g=2\pi/\lambda_g \approx 0.0014$ km$^{-1}$), where $\lambda_g$ is
the wavelength where $c_ik$ has a maximum. Of course, the Eady model
has been seminal because this wavelength is similar to the
wavelength of observed weather systems in the midlatitudes
\cite[see][for details]{Holton:2004,Pedlosky:1987,Gill:1982,Vallis:2006}.
The growth rate as a function of $k$ is shown 
in Fig.~\ref{Eady1} (solid line).

We can also calculate the poleward eddy heat flux of unstable Eady
waves (averaged over one wavelength $\lambda$):
\begin{equation}
\overline{v'T'}=\frac{1}{\lambda}\int_0^{\lambda} v'T' dx
= \frac{c_i \alpha^2 k f_0 H A^2  }{2 \Lambda R} \cos^2(ly)\exp^2(kc_it) \,\, ,
\label{heatflux-Eady}
\end{equation}
where $A$ is an (so far) arbitrary amplitude coefficient of the stream function
perturbation.  That is, $\overline{v'T'} \propto c_i \alpha^2 k$ from
which we can calculate the wavelength of maximum heat transport
$\lambda_h \approx 4200$ km (or $k_h=2\pi/\lambda_h \approx 0.0015$
km$^{-1}$), where $\lambda_h$ is the wavelength where $c_i \alpha^2 k$
has a maximum. The (dimensionless) heat transport as a
function of $k$ is shown in Fig.~\ref{Eady1} (doted line); it is 
arbitrarily scaled to have the same maximum value as the growth
rate to make the shape of the graphs easily comparable.
 
\begin{figure}[H]
\begin{center}
\includegraphics[scale=1.0,angle=0]{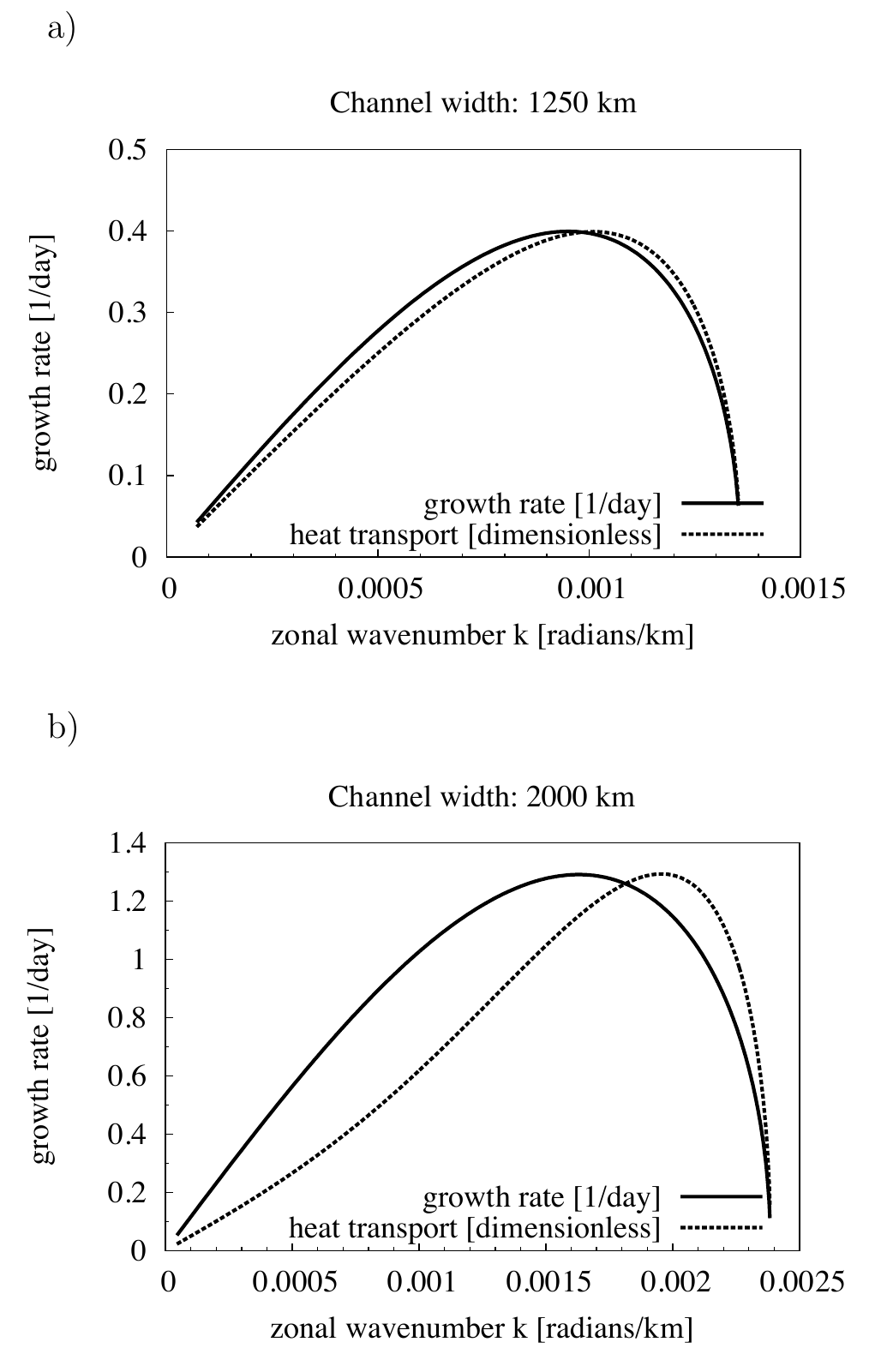}
\end{center}
\caption{\label{Eady2} Growth rate $c_ik$ (solid lines) and dimensionless 
eddy heat flux $\overline{v'T'} \propto c_i \alpha^2 k$ (dotted lines) of 
unstable Eady wave as a function of zonal wavenumber $k$ for
meridional channel width a) 1250 km, and b) 2000 km.}
\end{figure}

While the maxima $\lambda_g \approx 4500$ km and $\lambda_h \approx
4200$ km are not identical, they are very similar (i.e., have the same
order of magnitude). Thus, the maximum heat flux principle (derived
from MEP in our simplified setting) is consistent with the observed
wavelength of unstable baroclinic eddies in the midlatitude
atmosphere. In particular, the wavelength of the fasted growing wave
explains more than $95\%$ of the maximum heat transport. Different
channel width are explored in Fig.~\ref{Eady2}.  In Fig.~\ref{Eady2}a
the channel width is decreased to $L_y=$ 1250 km, resulting in almost
identical wavelengths of maximum entropy production (i.e., eddy heat
transport) and maximum growth rate. On the other hand, if the channel
width is increased to $L_y=$ 2000 km (Fig.~\ref{Eady2}b), the maxima
drift slightly further apart.  However, even in that case the
wavelength of the fasted growing wave explains more than $90\%$ of the
maximum heat transport.

\begin{figure}[H]
\begin{center}
\includegraphics[scale=1.0,angle=0]{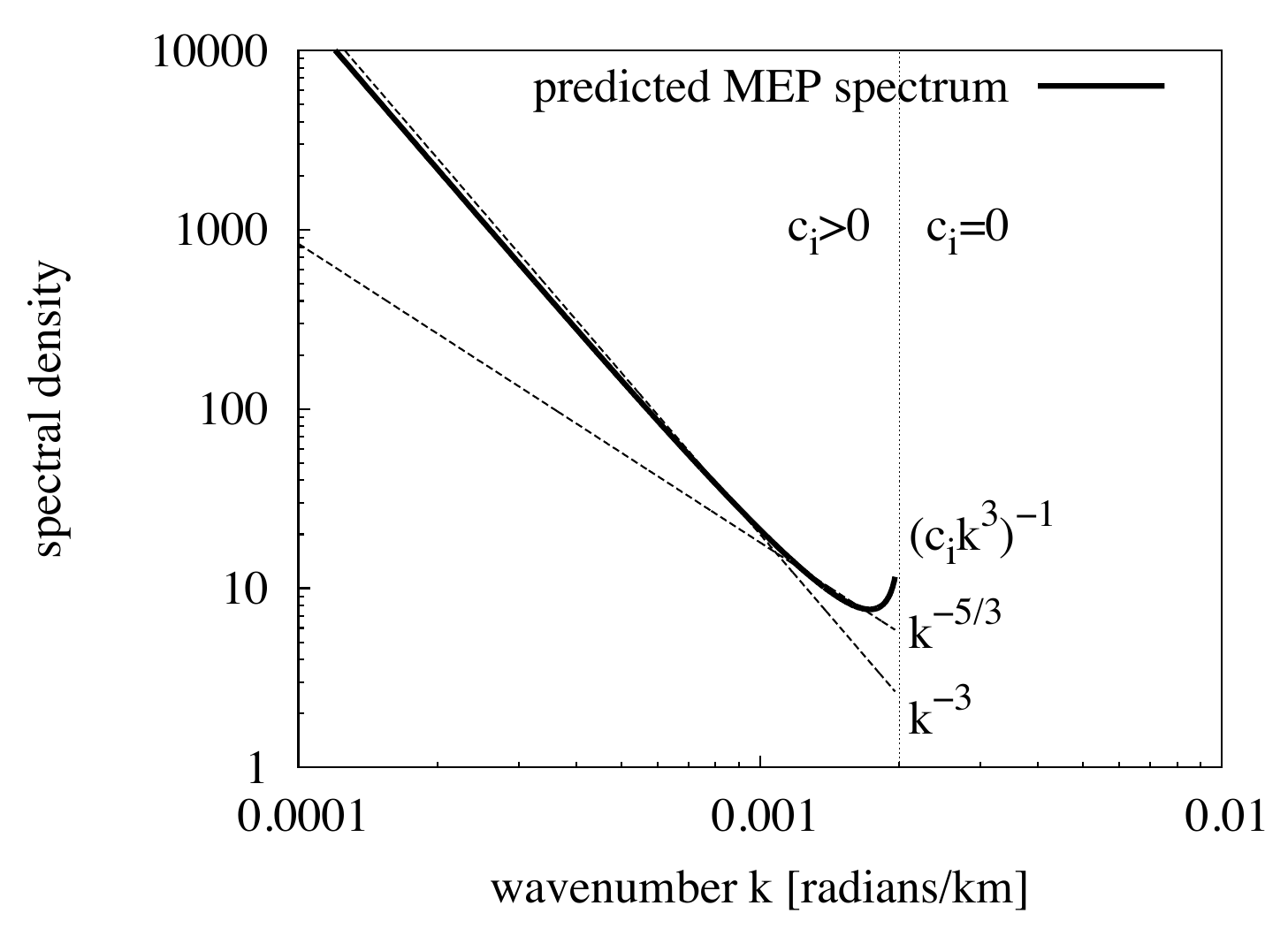}
\end{center}
\caption{\label{Spectrum-MEP} MEP spectrum $\propto ({c_ik^3})^{-1}$ of 
unstable Eady waves (solid line). Note that for small wavenumbers (long aves) the
spectrum approximately follows a $k^{-3}$ law (dashed line). For wavenumbers
larger than $k \approx 0.001$ radians/km the slope of the spectrum changes 
towards $k^{-5/3}$ (dashed line).}
\end{figure}

\begin{figure}[H]
\begin{center}
\includegraphics[scale=1.0,angle=0]{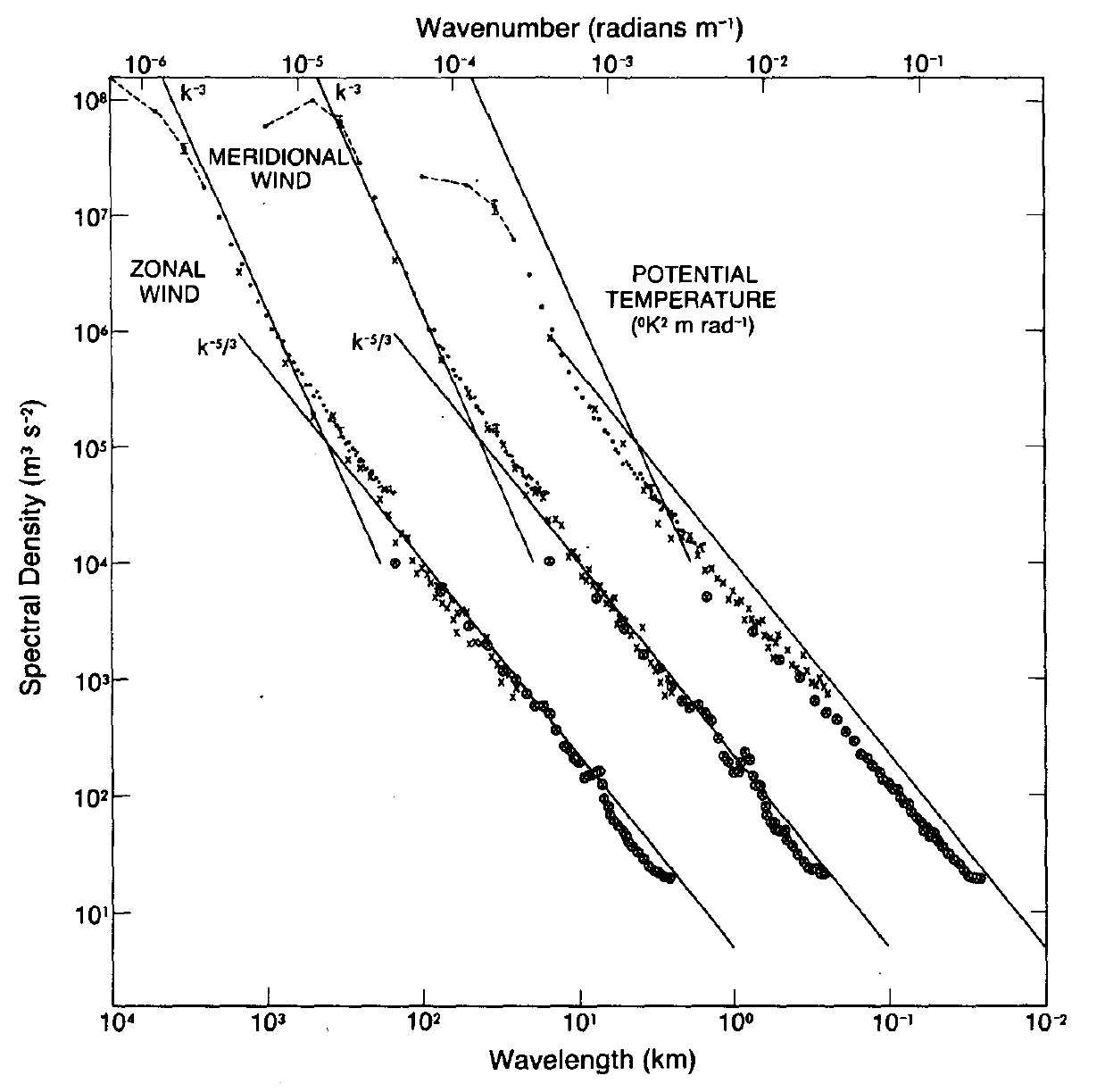}
\end{center}
\caption{\label{Spectrum-Observed}
Power spectra of wind and potential temperature near the tropopause
from GASP (Global Atmospheric Sampling Program) aircraft data. The
spectra for meridional wind and temperature are shifted one and two
decades to the right, respectively. Adapted from \cite{Nastrom:1985}.}
\end{figure}

Equation (\ref{heatflux-Eady}) describes the heat flux of a single
wave. However, fully developed baroclinic turbulence does consist of
a full spectrum of wavenumbers. Therefore, let us assume a turbulent
midlatitude atmosphere in which a full spectrum of baroclincally
growing eddies is already in place. We also assume that those eddies
are circular ($k=l$) and that each of them transports heat poleward
according to Eq.~(\ref{heatflux-Eady}). To obtain the total heat flux
$\overline{F}$ of a full spectrum of waves Eq.~(\ref{heatflux-Eady})
has to be summed/integrated over all wavenumbers $k$:
\begin{equation}
\overline{F} \propto \int_0^\infty A(k)^2c_ik^3 dk
= \textrm{Maximum} \,\, ,
\label{heatflux-Eady-2}
\end{equation}
where we skipped all the constants (including functions of $t$ and
$y$), used $\alpha^2=2k^2/\epsilon$ for circular eddies, and already
assumed that the total heat flux has to be maximized according to the
simplified version of the MEP principle. Note that now $A$ is not a
constant but a function of the wavenumber $k$: $A=A(k)$. Thus,
$A(k)^2$ is the spectral density in the wavenumber spectrum of the
stream function perturbation (and the related velocity and temperature
anomalies). That means Eq.~(\ref{heatflux-Eady-2}) asks us to choose
the wavenumber spectrum that maximizes the integral of all
wavenumbers. The maximum requires
\begin{equation}
\Delta \int_0^\infty A(k)^2c_ik^3dk = 
\int_0^\infty \Delta \left[A(k)^2c_ik^3 \right]dk = 0 \,\, ,
\end{equation}
or $A(k)^2c_ik^3 = \textrm{const}$. Therefore, the wavenumber spectrum
which maximizes the heat flux is
\begin{equation}
A(k)^2 \propto \frac{1}{c_ik^3} \,\, .
\label{spectrum}
\end{equation}
As we have $c_i$ from the dispersion relation, we can immediately
evaluate (\ref{spectrum}) (Fig.~\ref{Spectrum-MEP}).  It can be seen
that for long waves (small wavenumbers) the spectrum approximately
follows a $k^{-3}$ law. For wavelengths shorter than about 4000 km the
slope of the spectrum changes towards $k^{-5/3}$.  Close to the
instability limit (at $k \approx$ 0.002 radians/km or a wavelength of
about 3000 km) the spectrum bends upward. While we do not expect an
overly realistic spectrum from a model as simple as the Eady model,
the $-3$ slope for long waves is astonishingly accurate; an observed
atmospheric wavenumber spectrum is shown in
Fig.~\ref{Spectrum-Observed} [adapted from \cite{Nastrom:1985}].
There it can be seen that spectra of temperature and velocity have
slopes close to $-5/3$ for scales up to about 400 km. At larger scales
the spectra steepen to an approximate slope of $-3$. That is, we
conclude that the observed $k^{-3}$ spectrum for long waves is
consistent with heat flux maximization due to baroclinic instability
in the midlatitude atmosphere.

We are, of course, aware of the fact that the observed spectra in
Fig.~\ref{Spectrum-Observed} are traditionally explained through the
full nonlinear interactions in fully developed quasigeostrophic 2D and
3D turbulence, and that there exists an extensive pool of literature
(not referenced here) dealing with the spectra and the related energy
and enstrophy cascades. While there is no doubt that full nonlinear
models are the appropriate tools to study turbulence it is, however,
interesting that a simple linear model of baroclinic instability is
able to produce a $-3$ slope for long waves if we maximize the
meridional heat transport. A more detailed study of this phenomenon is
part of our ongoing research.
     
\subsubsection{Meridional heat transport and non-Gaussian variability}
\vspace{-3ex}

So far we have seen that heat flux maximization, derived from the MEP
principle, is capable of explaining some general features of observed
wavenumber scales/spectra in the midlatitude atmosphere. Let us now
explore if heat flux maximization is also consistent with the observed
non-Gaussian variability in the midlatitude free atmosphere. That is,
we ask if the midlatitude eddy heat flux $\overline{F}\propto
\overline{v'T'}$ is indeed maximized by non-Gaussian variability in
$v'$ and/or $T'$.

\begin{figure}[H]
\begin{center}
\includegraphics[scale=1.0,angle=0]{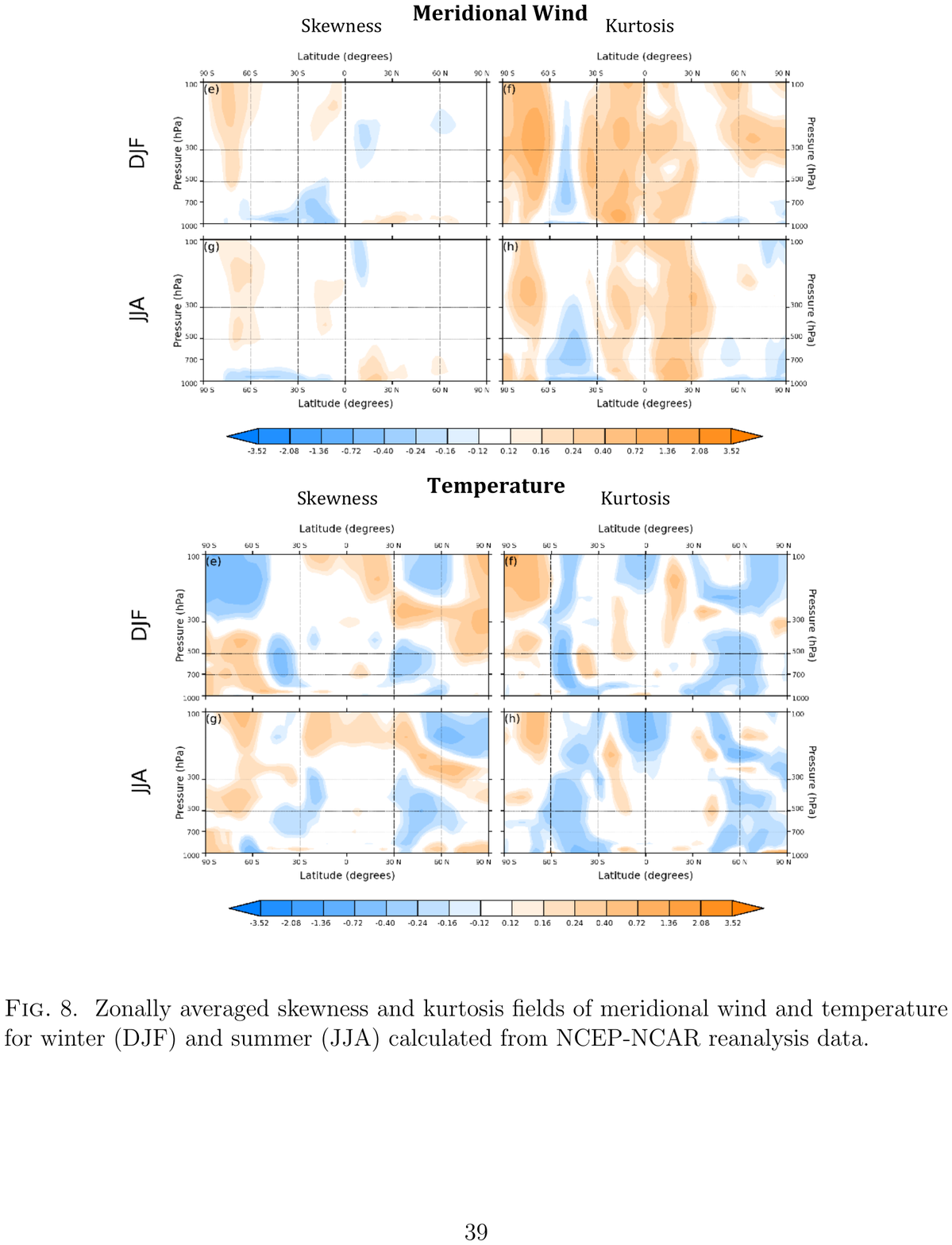}
\end{center}
\caption{\label{Moments}
Zonally averaged skewness and kurtosis
fields of meridional wind and temperature for winter (DJF) and summer
(JJA) calculated from NCEP-NCAR reanalysis data.}
\end{figure}

In general, the statistics of almost all atmospheric variables are
non-Gaussian. However, the origin of non-Gaussianity is not unique and
is a topic of active research [\cite{Sura:2014} provide a
comprehensive review, focusing on atmospheric synoptic and
low-frequency variability]. Often skewness $S$ and/or
excess\footnote{Here we use excess kurtosis, defined with respect to
  that of a Gaussian distribution.} kurtosis $K$ are used to quantify
non-Gaussianity: $S(x') \equiv x'^3/\sigma^3$ and $K(x') \equiv x'^4/\sigma^4
- 3$, for anomalies $x'$ with zero mean and standard deviation
$\sigma$.  For example, \cite{Perron:2013} provide a 62-year
(1948-2009) climatology of global skewness and (excess) kurtosis at
every gridpoint of key atmospheric variables (including meridional
wind and temperature anomalies $v'$ and $T'$, respectively) using
daily data from the NCEP-NCAR Reanalysis project
\cite[][]{Kalnay:1996}. The zonally averaged skewness and kurtosis
fields of meridional wind and temperature for winter (DJF) and summer
(JJA) are shown in Fig.~\ref{Moments} (however, in the following we
focus exclusively on the northern hemisphere winter).

\begin{figure}[H]
\begin{center}
\includegraphics[scale=1.0,angle=0]{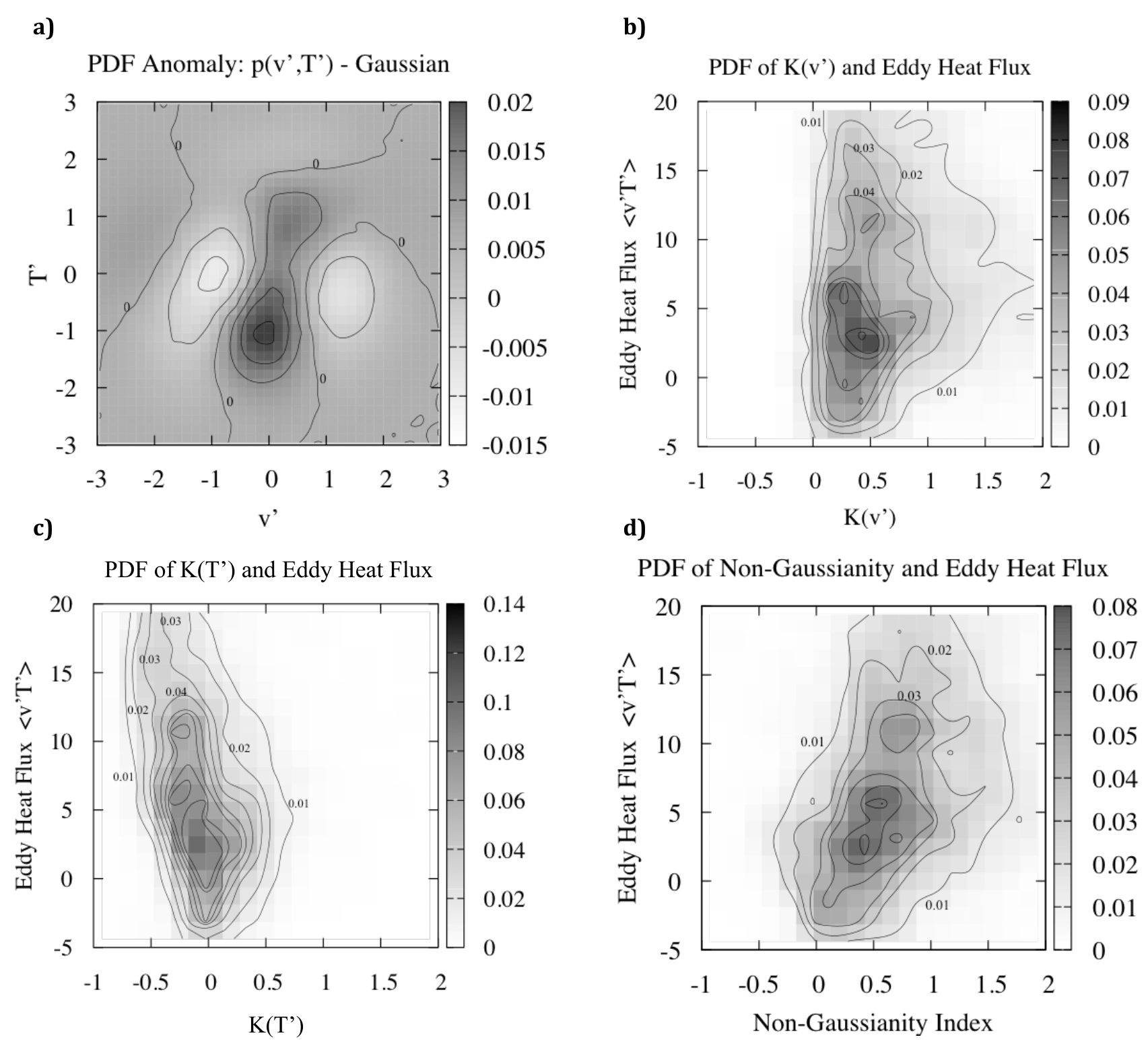}
\end{center}
\caption{\label{PDFs}
a) PDF anomaly [i.e., joint PDF $p(v',T')$ minus bivariate Gaussian]
of normalized midlatitude meridional wind $v'$ and temperature $T'$
anomalies.  The contour interval is 0.005.
b) Joint PDF $p(K(v'),\overline{v'T'})$. Note the positive correlation.
The contour interval is 0.01.
c) Joint PDF $p(K(T'),\overline{v'T'})$. Note the negative correlation.
The contour interval is 0.01.
d) Joint PDF $p(NGI,\overline{v'T'})$ of non-Gaussianity index
$NGI \equiv K(v') - K(T')$ and eddy heat flux $\overline{v'T'}$
for all midlatitude gridpoints. Note the positive correlation between the
non-Gaussianity index and the eddy heat flux. The contour interval is 0.01.}
\end{figure}

For our purpose, let us emphasis several points. The skewness of
meridional winds is close to zero in the entire midlatitude northern
hemisphere (in winter and summer). The kurtosis of meridional winds is
on average slightly positive in the midlatitude northern hemisphere
winter. The zonally averaged skewness and kurtosis fields of
temperature are a bit more complicated with respect to height and
latitude, but we can make the following general statements.  Averaged
over the entire midlatitude atmosphere the temperature skewness is
approximately zero. In addition, the temperature kurtosis is
predominantly (on average) negative in the midlatitude northern
hemisphere winter. Those general statements can also be visualized by
looking at the joint PDF of all midlatitude meridional wind and
temperature anomalies $v'$ and $T'$. To do that we normalize the
meridional wind and temperature time series at all midlatitude grid
points (approximately 25$^\circ$N -- 65$^\circ$N and 900 hPa -- 100
hPa) to zero mean and unit standard deviation. We then calculate the
joint PDF $p(v',T')$ and the related bivariate Gaussian distribution.
The non-Gaussian structure of $p(v',T')$ can now be highlighted by
plotting the joint PDF anomaly [i.e., $p(v',T')$ minus the bivariate
Gaussian]; the result is shown in Fig.~\ref{PDFs}a.
Of course, the first-order fact to notice is that $p(v',T')$ has a
pronounced non-Gaussian structure. There is a pronounced stronger than
Gaussian peak around about $v'\approx 0$ and $T'\approx - 1$. A
secondary peak exists near $v'\approx 0.3$ and $T'\approx + 1$. Both
peaks are connected by a stronger than Gaussian ridge.  On both side
of that ridge are weaker than Gaussian regions at about $v'\approx \pm
1$ and $T'\approx 0$. This overall structure results in a marginal
distribution $p(v')$ (not shown) with vanishing skewness and positive
kurtosis, and a marginal distribution $p(T')$ (not shown) with
negligible skewness and negative kurtosis, consistent with the above
discussion of Fig.~\ref{Moments} and the analysis of \cite{Messori:2012}.

Let us now analyze if the heat flux $\overline{F}\propto
\overline{v'T'}$ is maximized by non-Gaussian variability in $v'$
and/or $T'$. To do that we calculated the eddy heat flux
$\overline{v'T'}$ and the skewness and kurtosis of $v'$ and $T'$ at
every midlatitude gridpoint. We then looked at the joint PDF of the
individual moments [i.e., $S(v')$, $S(T')$, $K(v')$, $K(T')$] and
$\overline{v'T'}$ to find a potential correlation between
non-Gaussianity and the eddy heat flux.  It turns out that there is no
significant correlation between the skewness $K(v')$ and $K(T')$, and
$\overline{v'T'}$, which makes sense because $K(v')$ and $K(T')$ are
small (as discussed above). However, the eddy heat flux
$\overline{v'T'}$ is correlated with the kurtosis $K(v')$ and $K(T')$:
The correlation is {\em positive} between $\overline{v'T'}$ and
$K(v')$ (see joint PDF in Fig.~\ref{PDFs}b), and {\em negative} 
between $\overline{v'T'}$ and $K(T')$ (see joint PDF in
Fig.~\ref{PDFs}c).  To summarize (and visualize) this behavior we
constructed a non-Gaussianity index $NGI$ as $NGI \equiv K(v') -
K(T')$ (this index captures the observed correlation between the eddy
heat flux and the non-Gaussianity of $v'$ and $K'$) and calculated the
joint PDF of this index and the eddy heat flux. The result
$p(NGI,\overline{v'T'})$ is shown in Fig.~\ref{PDFs}d, where it can be
seen that there exists a (close to linear) positive correlation
between the non-Gaussianity index and the eddy heat flux. That is, a
positive $K(v')$ and a negative $K(T')$ are linked to a high
probability of a large positive (i.e., poleward) eddy heat flux
$\overline{v'T'}$. On the other hand, a close to Gaussian flow field
likely results in a relatively small eddy heat flux.

Overall, the midlatitude eddy heat flux $\overline{F}\propto
\overline{v'T'}$ is increased by non-Gaussian variability in $v'$ and
$T'$. In other words, the MEP derived maximum heat flux principle
principle is consistent with the observed non-Gaussian variability and
the related meridional eddy heat flux in the midlatitude free
atmosphere. While, at this point, we can only speculate (done in the
the following conclusions) about specific dynamical mechanisms linking
MEP and non-Gaussian variability, MEP (and related flux maximization
principles) seems a useful and promising tool to study and potentially
better understand non-Gaussian weather and climate variability.

\section{Conclusions}
\label{conclusions}
\begin{figure}[H]
\begin{center}
\includegraphics[scale=1.0,angle=0]{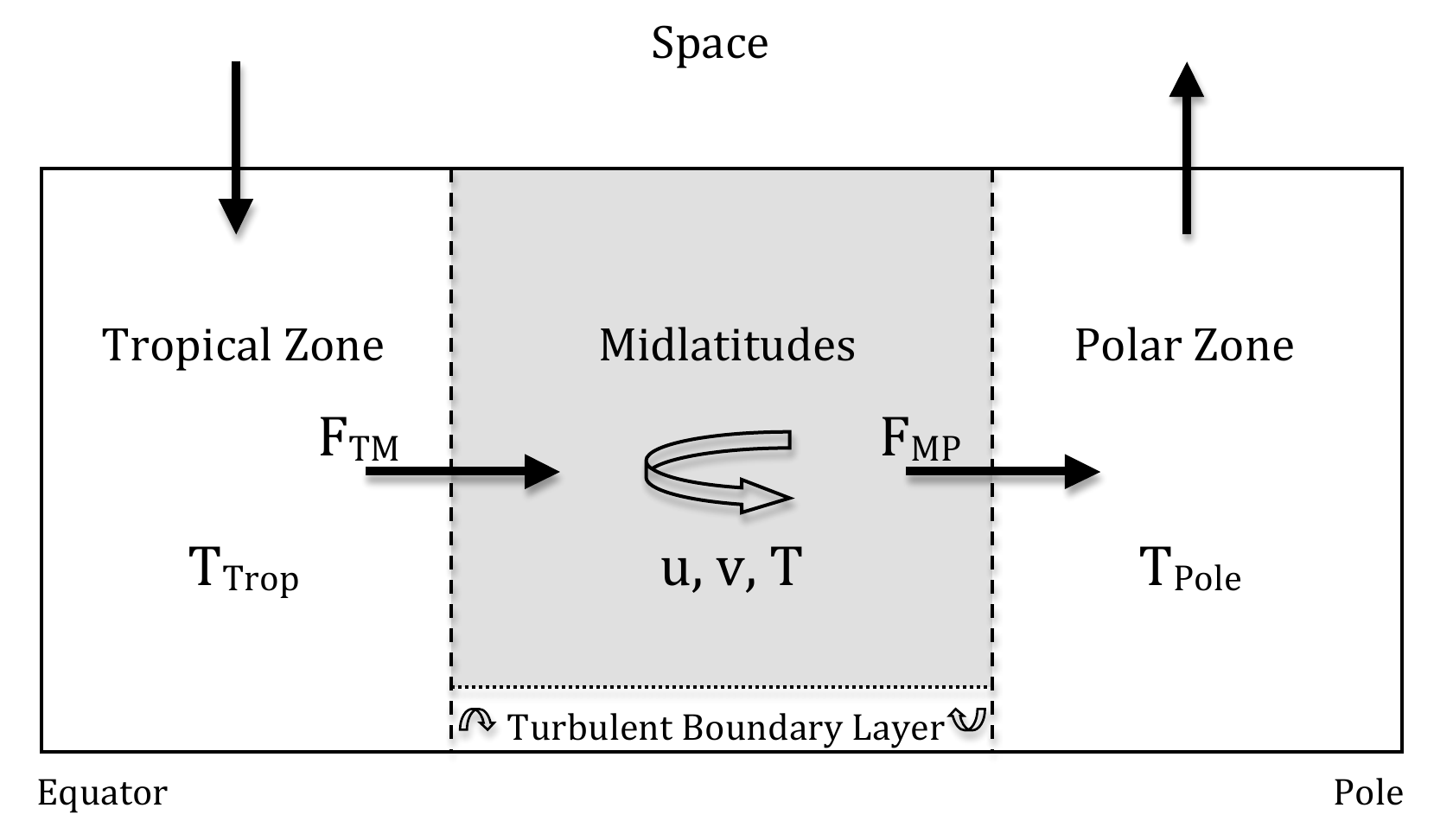}
\end{center}
\caption{\label{Figure-1} Representation of the atmosphere 
for the entropy production mechanisms considered in this paper.
The atmosphere is composed of three regions: a tropical zone 
(with temperature $T_{Trop}$), the midlatitudes (temperature $T$), 
and a polar zone (temperature $T_{Pole}$). In the tropics there is a 
net input of radiation/heat, in polar regions there is a net output. 
The midlatitude atmosphere transports heat poleward through its 
circulation (characterized by $u$, $v$, and $T$). The poleward heat 
flux consists of a flux from the tropics into the midlatitudes $F_{TM}$, 
and a flux from the midlatitudes into the polar zone $F_{MP}$.
Below the free midlatitude atmosphere we have a turbulent boundary 
layer that exerts a drag on the surface winds and induces a dissipative
heating rate $\Phi$. In this paper we consider the entropy
production through (a) viscous dissipation in the boundary layer, and (b)
the meridional heat flux in the midlatitude volume (the shaded region)
above the boundary layer.}
\end{figure}

In this paper we have presented theoretical and observational evidence
that some processes in the midlatitude atmosphere are significantly
non-Gaussian to maximize entropy production or related heat
fluxes. First we have shown that the skewness of sea surface winds,
damped by nonlinear surface drag, are consistent with maximum entropy
production (MEP). Then we considered a simplified conceptual model of
the atmosphere composed of three regions: a tropical zone, the
midlatitudes, and a polar zone (Fig.~\ref{Figure-1}). One novel
finding presented in this paper is that the observed $k^{-3}$
wavenumber spectrum of long waves in the midlatitudes can be roughly
explained by maximizing the meridional eddy heat flux
$\overline{v'T'}$ of unstable baroclinic Eady waves.  Finally we have
shown that the meridional eddy heat flux is increased by non-Gaussian
variability in meridional wind and temperatures anomalies $v'$ and
$T'$. Overall that means that MEP is consistent with the observed
non-Gaussian variability in the midlatitude atmosphere. That is, MEP
may point a way forward towards a unified perspective of
non-Gaussianity in weather and climate variability.

Why are our findings important and useful? One question often asked
with respect to climate change is the following: How does the
probability of extreme events changes in a warming climate? More often
than not an answer is provided by extensive numerical simulations.
Numerical modeling allows us to estimate the statistics of extreme
events (the tails of the PDF) by integrating a general circulation
model (GCM) for a very long period
\cite[e.g.,][]{Easterling:2000,Kharin:2005,Kharin:2007}. It is obvious
that the efforts by the Intergovernmental Panel on Climate Change
(IPCC) to understand and forecast the statistics of extreme weather
and climate events in a changing climate fall into this
category. However, while the numerical modeling is very useful and
practical, a weakness lies in the largely unknown ability of a GCM to
reproduce the correct statistics of extreme events.  Currently, many
climate models are calibrated to reproduce the observed first and
second moments (mean and variance) of the general circulation of the
ocean and atmosphere. Very little is known about the credibility of
GCMs to reproduce non-Gaussian statistics.

The MEP proposition allows us to provide a different, more
conceptually pleasing, kind of answer, though. The probabilities
(i.e., PDFs) of variables and processes under consideration are
altered by climate change to maximize entropy production. That is, by
changing the atmosphere's energy balance and related heat fluxes, the
climate system adjusts its PDFs to a new MEP state. While, at first
sight, this answer is not of much practical value compared to
straightforward GCM runs, the MEP principles potentially gives us a
tool to gauge the fidelity of global warming simulations in a
thermodynamically sound and well established framework.  The
exploration of this route is part of our ongoing and future research.

{\em Acknowledgments.}
We would like to thank Dr.~Hisashi Ozawa and one anonymous reviewer for
their constructive comments that helped to substantially improve the manuscript.


\clearpage
\newpage
\bibliographystyle{ametsoc} 
\bibliography{maxent}


\newpage
\renewcommand{\baselinestretch}{1.0}


\end{document}